\documentclass[pdflatex,sn-mathphys-num]{sn-jnl}

\usepackage{graphicx}%
\usepackage{multirow}%
\usepackage{amsmath,amssymb,amsfonts}%
\usepackage{amsthm}%
\usepackage{mathrsfs}%
\usepackage[title]{appendix}%
\usepackage{xcolor}%
\usepackage{textcomp}%
\usepackage{manyfoot}%
\usepackage{booktabs}%
\usepackage{algorithm}%
\usepackage{algorithmicx}%
\usepackage{algpseudocode}%
\usepackage{listings}%

\theoremstyle{thmstyleone}%
\theoremstyle{thmstyletwo}%

\theoremstyle{thmstylethree}%

\begin{document}

\title[Article Title]{Z(3) Metastable Bubbles and Chiral Dynamics Across a Dark-QCD Deconfinement Transition}


\author*[1,2]{\fnm{Jingxu} \sur{Wu}}\email{wuxj@my.msu.ru}

\author[1,2]{\fnm{Chenjia} \sur{Li}}\email{li.chenjia23@physics.msu.ru}
\equalcont{These authors contributed equally to this work.}

\author[1,2]{\fnm{Jie} \sur{Shi}}\email{shitcze@my.msu.ru}

\affil*[1]{\orgdiv{Faculty of Physics}, \orgname{Lomonosov Moscow State University}, \orgaddress{\city{Moscow}, \postcode{119991}, \country{Russia}}}

\affil[2]{\orgname{GeZhi International Theoretical Physics Reading Group}}


\abstract{
We present a self-contained theoretical analysis of a dark-QCD chiral transition in which the Polyakov-loop sector retains an explicit $Z(3)$ structure and couples consistently to the chiral order parameter.
Working within a coupled chiral--Polyakov effective theory, we map the homogeneous vacuum landscape and identify a metastability window bounded by spinodal loss of stability.
We then construct $Z(3)$ domain-wall solutions including chiral backreaction, extracting temperature-dependent wall profiles and surface tension.
Finally, we connect homogeneous metastability and wall microphysics to thermal bubble nucleation by evaluating the critical radius $R_c(T)$ and the nucleation exponent $S_3(T)/T$ in the thin-wall regime, providing a compact set of reproducible diagnostics for the decay of the metastable phase.
Our results establish a coherent pipeline from vacuum structure to nonperturbative interfaces and nucleation barriers, suitable for systematic extensions to full multi-field bounce calculations and dark-sector cosmological applications.
}


\maketitle

\section{Introduction}
\label{sec:intro}
Strongly coupled gauge theories can exhibit a qualitatively rich thermal phase structure, where confinement/deconfinement phenomena tied to center symmetry coexist and compete with the emergence (or disappearance) of chiral order. 
A standard diagnostic of confinement at finite temperature is the Polyakov loop, which in the heavy-quark (pure gauge) limit transforms nontrivially under the global center symmetry $Z(N_c)$ and serves as an order parameter for deconfinement \cite{McLerranSvetitsky1981,SvetitskyYaffe1982}. 
For $N_c=3$, the deconfined regime admits a characteristic $Z(3)$ pattern: three branches related by center rotations. 
With dynamical fundamental fermions the center symmetry is explicitly broken, the three branches are no longer exactly degenerate, yet metastable $Z(3)$-related extrema can persist over a finite temperature range above the crossover/transition region. 
Such metastability is intrinsically dynamical: it can end through non-equilibrium decay channels such as thermal bubble nucleation or spinodal-like instabilities \cite{Pisarski2000,Arefeva:2024wqb,Rusalev:2023imv,DumitruPisarski2001,Dumitru2002,Biswal2020}.

A complementary and equally important characterization of the thermal transition is provided by chiral order, for instance through an effective condensate $\sigma\sim\langle\bar\psi\psi\rangle$ in the light-quark regime. 
A central lesson from QCD-inspired modeling is that deconfinement and chiral restoration are generically entangled. 
The Polyakov-loop background modifies quark thermal excitations and thereby reshapes the chiral effective potential, while chiral dynamics feeds back into the effective strength of center-symmetry breaking and can shift the location and character of the deconfinement crossover/transition \cite{Fukushima2003,Ratti2006,Sasaki2007}. 
This interplay becomes especially consequential when one goes beyond homogeneous equilibrium and considers spatially non-uniform configurations. 
In $SU(3)$, the $Z(3)$ structure supports interfaces that interpolate between different center branches in the deconfined regime. 
Once the center symmetry is explicitly broken, such interfaces become biased: their energetics and stability are controlled by a competition between the vacuum free-energy splitting and the interface (domain-wall) tension. 
In a coupled description of chiral and Polyakov sectors, the chiral condensate generically varies across the interface, leading to a local suppression of $\sigma$---a form of spatially inhomogeneous chiral restoration---which in turn renormalizes the surface tension and can significantly alter nucleation barriers.

These considerations are particularly well motivated for dark confining sectors. 
Hidden $SU(3)_D$ (or more general $SU(N_c)_D$) gauge theories with light fermion content can support confinement and chiral transitions at scales unrelated to $\Lambda_{\rm QCD}$, and effective descriptions of their phase structure have been explored in multiple contexts \cite{Helmboldt2019,Reichert2022,Aoki2017}. 
In this work we deliberately adopt a purely theoretical perspective: rather than committing to a specific cosmological history or observational target, we focus on the universal field-theoretic ingredients that control $Z(3)$ metastability, interfaces, and metastable decay once chiral order is coupled consistently to the Polyakov sector.

Concretely, we construct a minimal three-dimensional finite-temperature Landau--Ginzburg description for a real chiral field $\sigma(\mathbf{x})$ and a complex Polyakov-loop field $\ell(\mathbf{x})$. 
The effective free energy includes gradient terms that penalize spatial variations, a chiral potential capturing restoration dynamics, a $Z(3)$-symmetric Polyakov potential responsible for the three-branch structure, an explicit breaking term induced by dynamical fermions, and an interaction term that locks the chiral and Polyakov sectors. 
Within this unified setup we aim to compute a coherent chain of quantities: 
 the homogeneous vacuum structure and the temperature window in which $Z(3)$-related branches remain metastable, 
 non-uniform interface solutions (domain walls) including chiral backreaction and their surface tension, and 
 critical bubbles mediating the decay of a metastable branch and the associated thermal nucleation suppression. 
The final step is naturally anchored in the semiclassical theory of metastable decay at finite temperature, where the dominant thermal saddle controls the exponential factor in the decay rate \cite{Langer1969,Linde1983,CallanColeman1977}.

The remainder of this paper is organized as follows. 
Section~2 specifies the effective potential terms, introduces a convenient dimensionless parametrization, and derives the coupled Euler--Lagrange equations for both homogeneous extrema and spatially varying configurations. 
Section~3 analyzes the homogeneous extrema, formulates a Hessian-based spinodal criterion for the disappearance of metastable branches, and extracts the temperature dependence of the vacuum splitting together with a practical metastability indicator. 
Section~4 constructs $Z(3)$ domain-wall solutions with chiral backreaction and computes the temperature dependence of the wall profiles, thickness, and surface tension. 
Section~5 turns to spherical critical bubbles and nucleation barriers, connecting the homogeneous vacuum splitting and interface tension to the temperature dependence of the critical radius and nucleation exponent, and discussing the crossover toward the instability-dominated regime near the spinodal edge. 
We conclude in Section~6 with a compact summary of universal trends and an outlook on systematic refinements, including full multi-field bounce computations, functional renormalization group improvements near criticality, and extensions to general $SU(N_c)_D$ and matter representations.

\section{Minimal coupled $(\sigma,\ell)$ effective theory and variational equations}
\label{sec:theory}

Our starting point is a three-dimensional, finite-temperature Landau--Ginzburg description for a real chiral order parameter $\sigma(\mathbf{x})\sim\langle\bar\psi\psi\rangle$ and a complex Polyakov-loop field $\ell(\mathbf{x})$ capturing the center-sector dynamics of an $SU(3)_D$ gauge theory. 
This approach follows the general philosophy of Wilson-line effective theories near deconfinement \cite{Pisarski2000,DumitruPisarski2001,Dumitru2002} and of chiral--Polyakov entanglement models used extensively in QCD thermodynamics \cite{Fukushima2003,Ratti2006,Sasaki2007,Biswal2020}. 
We emphasize that the present framework is meant to be \emph{minimal but dynamical}: it is sufficiently structured to support $Z(3)$ branches, metastability, and spatially varying interfaces/bubbles, while remaining analytically transparent.

We define the free-energy functional
\begin{align}
F[\sigma,\ell] \;=\; \int d^3x \Bigg\{&
\frac{K_\sigma}{2}\,(\nabla\sigma)^2
+\frac{K_\ell}{2}\,|\nabla\ell|^2
+ V_\chi(\sigma;T)
+ V_{Z(3)}(\ell,\bar\ell;T)
+ V_{\rm break}(\ell,\bar\ell;T)
+ V_{\rm int}(\sigma,\ell;T)
\Bigg\},
\label{eq:Ffunctional}
\end{align}
where $K_\sigma$ and $K_\ell$ are gradient (stiffness) parameters and the potentials are chosen to (i) realize chiral symmetry breaking/restoration in the $\sigma$ sector, (ii) realize the characteristic $Z(3)$ structure in the Polyakov sector, (iii) incorporate explicit center breaking due to dynamical fermions, and (iv) encode the backreaction between the two sectors.

For the chiral sector we adopt the standard quartic Landau form with explicit breaking,
\begin{equation}
V_\chi(\sigma;T)=\frac{a(T)}{2}\,\sigma^2+\frac{\lambda}{4}\,\sigma^4 - H\,\sigma,
\qquad
a(T)=a_0\left(\frac{T-T_\chi^0}{T_\chi^0}\right),
\label{eq:Vchi}
\end{equation}
where $H$ parametrizes explicit chiral symmetry breaking (finite fermion mass), while $a_0,\lambda,T_\chi^0$ control the scale and nature of the chiral transition in the absence of coupling to $\ell$. 
More refined chiral potentials (e.g.\ quark--meson or NJL mean-field reductions) can be mapped onto \eqref{eq:Vchi} by a local expansion in $\sigma$ around the transition region; our focus is on universal interface and metastability physics rather than on a detailed microscopic matching.

For the Polyakov sector we use the simplest polynomial $Z(3)$-invariant potential containing the cubic invariant $(\ell^3+\bar\ell^3)$,
\begin{equation}
V_{Z(3)}(\ell,\bar\ell;T)=
-\frac{b_2(T)}{2}\,|\ell|^2
-\frac{b_3}{6}\,(\ell^3+\bar\ell^3)
+\frac{b_4}{4}\,|\ell|^4,
\label{eq:VZ3}
\end{equation}
which is the canonical form that produces three deconfined branches at $N_c=3$ and has been widely used in PNJL/PQM modeling \cite{Ratti2006,Sasaki2007,Biswal2020}. 
Writing $\ell=\rho e^{i\theta}$ makes the $Z(3)$ structure explicit:
\begin{equation}
\ell^3+\bar\ell^3 = 2\rho^3\cos(3\theta),
\label{eq:Z3cos}
\end{equation}
so that for vanishing explicit breaking the deconfined minima occur at phases $\theta=0,2\pi/3,4\pi/3$ when $b_2(T)$ is sufficiently positive. 
The temperature dependence of $b_2(T)$ is taken as a smooth polynomial in $T$ (or in $T_0/T$) in the standard way, chosen such that $\ell=0$ is favored at low $T$ and $|\ell|>0$ at high $T$ \cite{Ratti2006}. 
Since the present paper targets structural rather than quantitative matching, we keep $b_2(T)$ generic while imposing only these qualitative constraints.

Dynamical fundamental fermions explicitly break center symmetry; in the Polyakov effective theory this is commonly represented at leading order by a linear ``tilt'' term \cite{Fukushima2003,Sasaki2007,Biswal2020},
\begin{equation}
V_{\rm break}(\ell,\bar\ell;T) = -h(T)\,(\ell+\bar\ell),
\label{eq:Vbreak}
\end{equation}
which lifts the degeneracy among the three $Z(3)$-related branches and turns two of them into metastable extrema over an intermediate temperature interval. 
The function $h(T)$ summarizes the strength of explicit breaking (controlled, in microscopic treatments, by the fermion content and masses); here it will be treated as a tunable parameter controlling the size of the metastability window.

Finally, the essential chiral--Polyakov backreaction is encoded in the simplest symmetry-allowed coupling,
\begin{equation}
V_{\rm int}(\sigma,\ell;T) = - g\,\sigma^2|\ell|^2 ,
\label{eq:Vint}
\end{equation}
which captures the generic tendency toward ``locking'': when $|\ell|$ grows (deconfinement tendency), the effective mass term for $\sigma$ is shifted and chiral restoration is promoted, while a large $\sigma$ suppresses $|\ell|$ through the same coupling. 
This quadratic coupling is the minimal choice compatible with $\sigma\to-\sigma$ symmetry when $H=0$ and is sufficient to generate pronounced chiral backreaction on interface/bubble profiles, which is the central theme of this work. 
Subleading couplings (e.g.\ $-\tilde g\,\sigma(\ell+\bar\ell)$) can be included to test robustness, but we will not need them for the qualitative phenomena discussed below.

The equilibrium configurations follow from the variational conditions $\delta F/\delta\sigma=0$ and $\delta F/\delta\bar\ell=0$,
\begin{align}
-K_\sigma \nabla^2\sigma(\mathbf{x}) + \frac{\partial V}{\partial\sigma}(\sigma,\ell,\bar\ell;T) &= 0,
\label{eq:EOMsigma}\\
-K_\ell \nabla^2\ell(\mathbf{x}) + \frac{\partial V}{\partial\bar\ell}(\sigma,\ell,\bar\ell;T) &= 0,
\label{eq:EOMell}
\end{align}
with the total local potential
\begin{equation}
V(\sigma,\ell,\bar\ell;T) \equiv V_\chi(\sigma;T)+V_{Z(3)}(\ell,\bar\ell;T)+V_{\rm break}(\ell,\bar\ell;T)+V_{\rm int}(\sigma,\ell;T).
\label{eq:Vtotal}
\end{equation}
For homogeneous phases, Eqs.~\eqref{eq:EOMsigma}--\eqref{eq:EOMell} reduce to algebraic stationarity conditions,
\begin{equation}
\frac{\partial V}{\partial\sigma}=0,\qquad
\frac{\partial V}{\partial\ell}=0,\qquad
\frac{\partial V}{\partial\bar\ell}=0,
\label{eq:gap}
\end{equation}
which determine the true vacuum and the competing metastable extrema (Sec.~\ref{sec:homog}). 
For spatially varying interfaces (domain walls) and critical bubbles, the full partial differential equations \eqref{eq:EOMsigma}--\eqref{eq:EOMell} will be solved under the appropriate boundary conditions; the resulting profiles and the derived surface tension and nucleation action form the core observables in Secs.~\ref{sec:domainwalls}--\ref{sec:bubbles}.

It is convenient to reduce the parameter dependence by an explicit nondimensionalization. 
Let $\sigma_0$ denote the characteristic chiral scale (e.g.\ the $T=0$ minimum at $H\to 0$) and $\ell_0$ the characteristic deconfined magnitude (e.g.\ the minimum of $V_{Z(3)}$ at high $T$ and $h\to0$). 
Introduce dimensionless fields and coordinates,
\begin{equation}
\sigma=\sigma_0 \hat\sigma,\qquad
\ell=\ell_0 \hat\ell,\qquad
\mathbf{x}=\xi\,\hat{\mathbf{x}},
\label{eq:rescale}
\end{equation}
with $\xi$ chosen such that one gradient coefficient is set to unity (e.g.\ $\xi^2=K_\sigma/a_0$ when $a_0>0$). 
After this rescaling, the model is governed by a small set of dimensionless couplings, for instance
\begin{equation}
\hat g = \frac{g\,\sigma_0^2}{b_4\,\ell_0^2},\qquad
\hat h = \frac{h}{b_4\,\ell_0^3},\qquad
\kappa = \frac{K_\sigma\,\sigma_0^2}{K_\ell\,\ell_0^2},
\label{eq:dimensionless}
\end{equation}
together with dimensionless versions of $a(T)$ and $b_2(T)$. 
This parametrization will be used to present results in a form that does not depend on arbitrary choices of units and highlights which combinations control metastability, interface thickness, and nucleation barriers.

\begin{figure}[htp]
  \centering
  \includegraphics[width=0.9\textwidth]{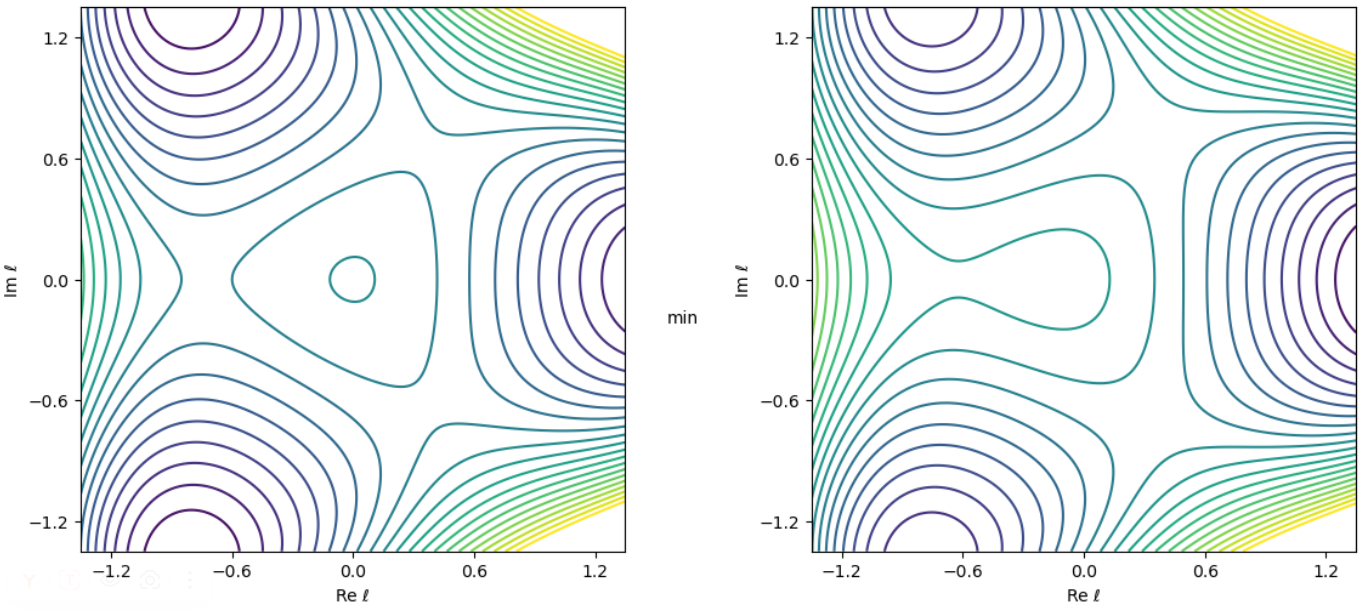}
  \caption{Schematic visualization of the Polyakov-loop sector potential: the cubic invariant in \eqref{eq:VZ3} generates the $Z(3)$ three-branch structure, while the explicit breaking term \eqref{eq:Vbreak} biases the branches and enables metastability. (Actual plots will be provided in Sec.~\ref{sec:homog} once parameters are fixed.)}
  \label{fig:Vell}
\end{figure}

\section{Homogeneous vacuum structure, metastability window, and spinodal criterion}
\label{sec:homog}

We first determine the homogeneous extrema of the coupled $(\sigma,\ell)$ theory and extract the quantities that will control interface and bubble dynamics in later sections. 
For spatially uniform configurations the gradient terms in \eqref{eq:Ffunctional} drop out, and equilibrium states are stationary points of the local potential
\begin{equation}
V(\sigma,\ell,\bar\ell;T)\equiv 
V_\chi(\sigma;T)+V_{Z(3)}(\ell,\bar\ell;T)+V_{\rm break}(\ell,\bar\ell;T)+V_{\rm int}(\sigma,\ell;T).
\end{equation}
The stationarity (gap) conditions are
\begin{equation}
\frac{\partial V}{\partial\sigma}=0,\qquad
\frac{\partial V}{\partial\ell}=0,\qquad
\frac{\partial V}{\partial\bar\ell}=0,
\label{eq:gap_homog}
\end{equation}
or, equivalently, in real variables $\ell\equiv \ell_R+i\ell_I$,
\begin{equation}
\frac{\partial V}{\partial\sigma}=0,\qquad
\frac{\partial V}{\partial \ell_R}=0,\qquad
\frac{\partial V}{\partial \ell_I}=0.
\end{equation}
Using \eqref{eq:Vchi}--\eqref{eq:Vint}, these equations read explicitly
\begin{align}
a(T)\,\sigma+\lambda\sigma^3-H-2g\,\sigma|\ell|^2 &= 0,
\label{eq:gap_sigma}\\
-b_2(T)\,\ell_R - b_3(\ell_R^2-\ell_I^2) + b_4|\ell|^2\ell_R -2h(T) -2g\,\sigma^2\ell_R &= 0,
\label{eq:gap_lR}\\
-b_2(T)\,\ell_I + 2b_3\,\ell_R\ell_I + b_4|\ell|^2\ell_I -2g\,\sigma^2\ell_I &= 0.
\label{eq:gap_lI}
\end{align}
The $Z(3)$ structure is most transparent in polar form $\ell=\rho e^{i\theta}$, where the cubic invariant becomes $\ell^3+\bar\ell^3=2\rho^3\cos(3\theta)$, implying three symmetry-related deconfined branches at $\theta=0,2\pi/3,4\pi/3$ when $h(T)=0$ and the deconfined magnitude $\rho$ is favored. 
Once $h(T)\neq 0$, the linear tilt selects a unique true branch (closest to $\theta=0$ for $h>0$) while the other two survive, if at all, as metastable extrema over an intermediate temperature window. 
For each $T$ we therefore identify the set of homogeneous extrema of $V$ and classify them as the true vacuum (global minimum), metastable vacua (local minima), and saddles (unstable stationary points). 
This produces the temperature-dependent homogeneous order parameters
\begin{equation}
(\sigma_{\rm true}(T),\ell_{\rm true}(T)),\qquad
(\sigma_{\rm meta}(T),\ell_{\rm meta}(T)),
\end{equation}
which serve as boundary data for the domain-wall and bubble problems.

A central quantity controlling the decay of a metastable branch is the vacuum splitting
\begin{equation}
\Delta V(T) \;\equiv\; V_{\rm meta}(T)-V_{\rm true}(T)\,>\,0,
\label{eq:DeltaV_def}
\end{equation}
evaluated at the corresponding homogeneous minima. 
In the thin-wall regime, $\Delta V(T)$ provides the bulk driving force favoring the true vacuum and enters the critical bubble radius and nucleation action, while beyond thin-wall it remains a faithful measure of the energetic bias between branches. 
The metastability \emph{window} is defined as the set of temperatures for which the metastable extrema exist as genuine local minima, i.e.\ have positive curvature in all independent directions in field space. 
Its endpoints are spinodals: beyond them, the metastable minimum merges with a saddle in a saddle--node bifurcation and ceases to exist.

This statement can be made precise through the Hessian matrix of the local potential at a stationary point. 
Let $\Phi\equiv(\sigma,\ell_R,\ell_I)$ and consider homogeneous fluctuations around a stationary point $\Phi_\star(T)$. 
The Hessian is
\begin{equation}
\mathcal{H}_{ij}(T)\;\equiv\;\left.\frac{\partial^2 V}{\partial \Phi_i\,\partial \Phi_j}\right|_{\Phi=\Phi_\star(T)},
\qquad i,j\in\{\sigma,\ell_R,\ell_I\}.
\label{eq:Hessian_def}
\end{equation}
A stationary point is a local minimum iff $\mathcal{H}$ is positive definite, i.e.\ all eigenvalues are strictly positive. 
We therefore define the stability diagnostic
\begin{equation}
\lambda_{\min}(T)\equiv \min \mathrm{eig}\,\mathcal{H}(T),
\label{eq:lmin_def}
\end{equation}
and identify the spinodal endpoint of metastability by
\begin{equation}
\lambda_{\min}(T)\;=\;0,
\label{eq:spinodal}
\end{equation}
evaluated at the metastable branch $\Phi_\star=\Phi_{\rm meta}(T)$. 
Equation \eqref{eq:spinodal} provides an unambiguous and coordinate-independent definition of the metastability-disappearance line in parameter space (e.g.\ in the $(T,h)$ plane, or equivalently in dimensionless variables $(\hat T,\hat h)$). 
For the explicit potentials \eqref{eq:Vchi}--\eqref{eq:Vint}, the Hessian entries can be written in closed form. 
Defining $\rho^2\equiv|\ell|^2=\ell_R^2+\ell_I^2$ and evaluating at a stationary point, one finds
\begin{align}
\frac{\partial^2 V}{\partial \sigma^2} &= a(T)+3\lambda\sigma^2-2g\rho^2,
\label{eq:Hss}\\
\frac{\partial^2 V}{\partial \sigma\,\partial \ell_R} &= -4g\,\sigma\,\ell_R,
\qquad
\frac{\partial^2 V}{\partial \sigma\,\partial \ell_I} = -4g\,\sigma\,\ell_I,
\label{eq:Hsl}\\
\frac{\partial^2 V}{\partial \ell_R^2} &= -b_2(T)-2b_3\ell_R + b_4(\rho^2+2\ell_R^2) -2g\sigma^2,
\label{eq:Hrr}\\
\frac{\partial^2 V}{\partial \ell_I^2} &= -b_2(T)+2b_3\ell_R + b_4(\rho^2+2\ell_I^2) -2g\sigma^2,
\label{eq:Hii}\\
\frac{\partial^2 V}{\partial \ell_R\,\partial \ell_I} &= 2\ell_I\,(b_4\ell_R+b_3).
\label{eq:Hri}
\end{align}
These expressions show explicitly how the entanglement coupling $g$ ties the curvature in the chiral direction to the Polyakov-loop directions. 
In particular, the same coupling that ``locks'' the homogeneous order parameters also shifts the approach to spinodality and hence changes the width of the metastability window. 
In later sections, the homogeneous quantities extracted here will enter the non-uniform solutions in two complementary ways: the values $(\sigma_{\rm true},\ell_{\rm true})$ and $(\sigma_{\rm meta},\ell_{\rm meta})$ provide boundary conditions for domain walls and critical bubbles, while $\Delta V(T)$ provides the bulk bias competing against the surface tension.

For reference, the main outputs of this section are the temperature-dependent homogeneous minima and their energetic splitting,
\begin{equation}
(\sigma_{\rm true}(T),\ell_{\rm true}(T)),\quad
(\sigma_{\rm meta}(T),\ell_{\rm meta}(T)),\quad
\Delta V(T)=V_{\rm meta}(T)-V_{\rm true}(T),
\end{equation}
together with the spinodal criterion \eqref{eq:spinodal} defined via the smallest Hessian eigenvalue at the metastable minimum. 
In Sec.~\ref{sec:domainwalls} we will compute the domain-wall profiles and the surface tension $\Sigma(T)$, while in Sec.~\ref{sec:bubbles} the pair $\{\Delta V(T),\Sigma(T)\}$ will control the critical bubble radius and the nucleation action.

\begin{figure}[htp]
  \centering
  \includegraphics[width=0.9\textwidth]{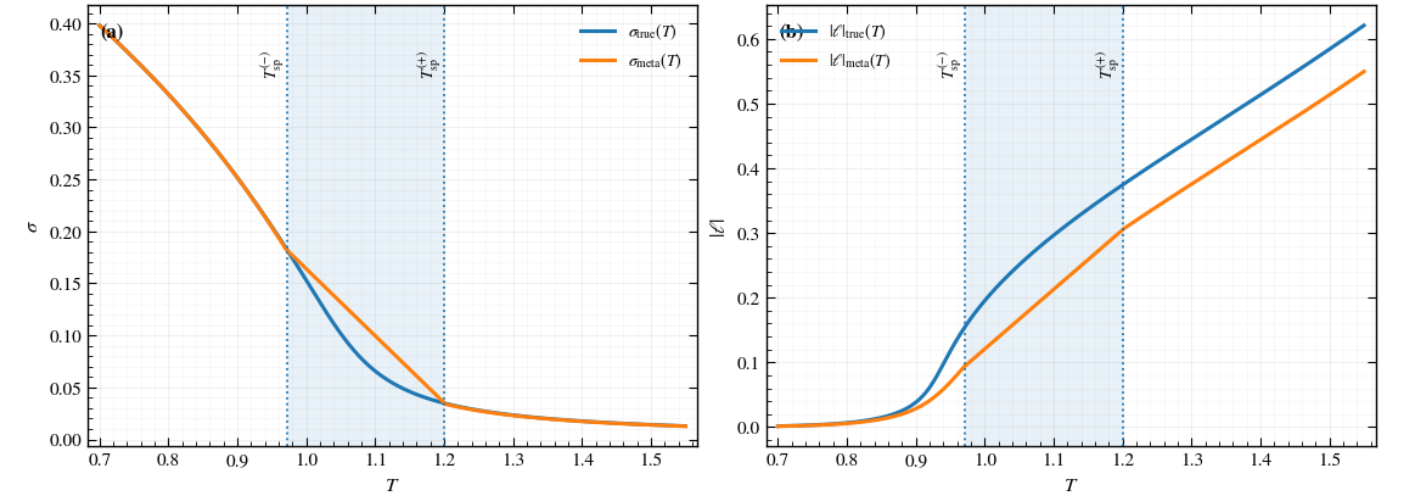}
  \caption{Temperature dependence of homogeneous order parameters in the coupled $(\sigma,\ell)$ theory.}
  \label{fig:orderparamsT}
\end{figure}

\begin{figure}[htp]
  \centering
  \includegraphics[width=0.9\textwidth]{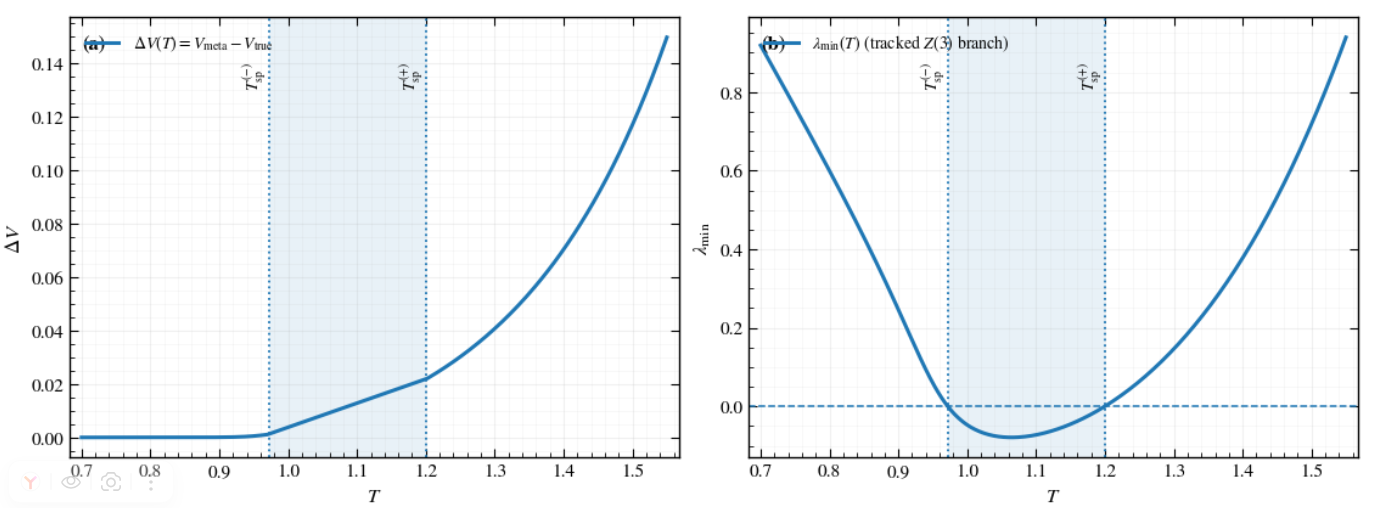}
  \caption{Vacuum splitting and spinodal criterion. The metastability window ends when the smallest Hessian eigenvalue at the metastable minimum vanishes.}
  \label{fig:DeltaV_spinodal}
\end{figure}

\section{Z(3) Domain Walls with Chiral Backreaction: Profiles and Surface Tension}
\label{sec:domainwalls}

Within the metastability window established in Sec.~3, the effective potential admits multiple (meta)stable homogeneous vacua whose Polyakov-loop phases differ by approximately $2\pi/3$ in the near--$Z(3)$ regime. 
A spatial interface interpolating between two such vacua is a $Z(3)$ domain wall. 
In the present theory this interface is intrinsically \emph{mixed}: gradients and phase rotation of the complex Polyakov loop induce a nontrivial chiral response through the coupling $-g\sigma^2|\ell|^2$, while the chiral profile backreacts on the Polyakov-loop wall by renormalizing the effective curvature and barrier along the amplitude direction. 
This section formulates the planar wall as a one-dimensional saddle-point problem, derives the profile equations and a useful first integral, and gives controlled expressions for the wall thickness and surface tension. 
Figures are reserved as placeholders at the end of the section.

We consider a static planar wall with normal direction $z$, so that $\sigma=\sigma(z)$ and $\ell=\ell(z)$ only. 
The free-energy functional per unit transverse area is taken as a minimal Landau--Ginzburg form,
\begin{equation}
\mathcal{F}[\sigma,\ell]
=\int_{-\infty}^{+\infty}\!dz\;
\left[
\frac{\kappa_\sigma}{2}\big(\partial_z\sigma\big)^2
+\kappa_\ell\,\big|\partial_z\ell\big|^2
+V_{\rm eff}\!\big(\sigma,\ell;T\big)
\right],
\label{eq:F_wall}
\end{equation}
where $\kappa_\sigma$ and $\kappa_\ell$ are stiffness coefficients and $V_{\rm eff}$ is the homogeneous effective potential defined in Sec.~3. 
For concreteness, and to maintain continuity with the numerical figures in Sec.~3, we recall the minimal decomposition
\begin{equation}
V_{\rm eff}(\sigma,\ell;T)=V_\chi(\sigma;T)+V_{Z(3)}(\ell;T)+V_{\rm br}(\ell;T)+V_{\rm int}(\sigma,\ell;T),
\end{equation}
with
\begin{align}
V_\chi(\sigma;T)&=\frac{a(T)}{2}\sigma^2+\frac{\lambda}{4}\sigma^4-H\,\sigma, \\
V_{Z(3)}(\ell;T)&=-\frac{b_2(T)}{2}|\ell|^2-\frac{b_3}{3}\Re(\ell^3)+\frac{b_4}{4}|\ell|^4, \\
V_{\rm br}(\ell;T)&=-h(T)\big(\ell+\ell^\ast\big)=-2h(T)\Re\ell, \\
V_{\rm int}(\sigma,\ell;T)&=-g\,\sigma^2|\ell|^2 .
\end{align}
The cubic anisotropy $\Re(\ell^3)$ encodes the $Z(3)$ structure: when $h(T)\to 0$ the three orientations $\arg\ell=0,\pm 2\pi/3$ are symmetry-related, while $h(T)\neq0$ selects a true vacuum and renders the other orientations metastable for intermediate temperatures. 
The coupling $g>0$ realizes the PNJL-type feedback: increasing $|\ell|$ shifts the effective chiral curvature $a(T)\mapsto a(T)-2g|\ell|^2$ and thus reduces $\sigma$ locally, whereas a reduced $\sigma$ weakens the $|\ell|$-mass term, thereby reshaping the wall.

The domain wall is specified by boundary conditions at spatial infinity,
\begin{equation}
(\sigma,\ell)\to(\sigma_-,\ell_-)\quad (z\to-\infty),\qquad
(\sigma,\ell)\to(\sigma_+,\ell_+)\quad (z\to+\infty),
\label{eq:bc_wall}
\end{equation}
where $(\sigma_\pm,\ell_\pm)$ are stationary points of $V_{\rm eff}$ at fixed $T$:
\begin{equation}
\partial_\sigma V_{\rm eff}(\sigma_\pm,\ell_\pm;T)=0,\qquad
\partial_{\ell^\ast} V_{\rm eff}(\sigma_\pm,\ell_\pm;T)=0.
\end{equation}
In the ideal $Z(3)$ limit one typically considers walls connecting \emph{degenerate} vacua. 
In the metastable regime, however, the two endpoints have slightly different vacuum energies, $\Delta V(T)=V_{\rm eff}(\sigma_+,\ell_+;T)-V_{\rm eff}(\sigma_-,\ell_-;T)\neq0$. 
A perfectly static planar wall in infinite volume is then not an exact equilibrium configuration because the pressure difference drives motion. 
Nevertheless, as long as $|\Delta V|$ is small compared to the barrier height separating the vacua (which is precisely the regime in the interior of the metastability window), the interface remains a well-defined quasi-static object whose intrinsic profile and surface tension control the thin-wall nucleation physics. 
In this section we therefore compute the stationary wall profile at fixed $T$ with \eqref{eq:bc_wall} and interpret the resulting $\Sigma(T)$ as the intrinsic tension entering the nucleation action.

Varying \eqref{eq:F_wall} yields coupled Euler--Lagrange equations,
\begin{align}
-\kappa_\sigma\,\sigma''(z)+\frac{\partial V_{\rm eff}}{\partial \sigma}\big(\sigma,\ell;T\big)&=0,
\label{eq:EOMsig_wall}\\
-\kappa_\ell\,\ell''(z)+\frac{\partial V_{\rm eff}}{\partial \ell^\ast}\big(\sigma,\ell;T\big)&=0,
\label{eq:EOMell_wall}
\end{align}
with
\begin{align}
\frac{\partial V_{\rm eff}}{\partial \sigma}
&=a(T)\sigma+\lambda\sigma^3-H-2g\sigma|\ell|^2,
\label{eq:dVds_wall}\\
\frac{\partial V_{\rm eff}}{\partial \ell^\ast}
&=-\frac{b_2(T)}{2}\ell-\frac{b_3}{2}\ell^{\ast\,2}+\frac{b_4}{2}|\ell|^2\ell-h(T)-g\sigma^2\ell.
\label{eq:dVdl_wall}
\end{align}
These equations are best viewed as a three-component real system by writing $\ell=x+iy$, but their complex form is compact and emphasizes that the backreaction enters through the $\sigma^2\ell$ term and through $|\ell|^2$ in the chiral force. 

A crucial structural identity follows from translational invariance in $z$. 
Multiplying \eqref{eq:EOMsig_wall} by $\sigma'$ and \eqref{eq:EOMell_wall} by $\ell'^\ast$ (plus complex conjugate) gives the ``energy conservation'' first integral
\begin{equation}
\frac{\kappa_\sigma}{2}\sigma'^2+\kappa_\ell|\ell'|^2
-V_{\rm eff}(\sigma,\ell;T)
=\text{const.}
\label{eq:firstint_wall}
\end{equation}
Evaluating at $z\to\pm\infty$ yields
\begin{equation}
\frac{\kappa_\sigma}{2}\sigma'^2+\kappa_\ell|\ell'|^2
=
V_{\rm eff}(\sigma,\ell;T)-V_{\rm eff}(\sigma_{\rm ref},\ell_{\rm ref};T),
\label{eq:firstint_ref}
\end{equation}
where $(\sigma_{\rm ref},\ell_{\rm ref})$ is the chosen reference vacuum used to fix the constant. 
For degenerate walls one can take either endpoint; for metastable walls we take the \emph{true} vacuum (lower energy) as the reference. 
Equation \eqref{eq:firstint_ref} is valuable numerically: it provides a stringent diagnostic for profile convergence and permits rewriting the surface tension as a manifestly positive integral in the near-degenerate regime.

To make the $Z(3)$ geometry explicit it is convenient to use polar variables for the Polyakov loop,
\begin{equation}
\ell(z)=\rho(z)e^{i\theta(z)},\qquad \rho\ge0,
\label{eq:ell_polar_wall}
\end{equation}
so that $|\ell'|^2=(\rho')^2+\rho^2(\theta')^2$ and $\Re(\ell^3)=\rho^3\cos(3\theta)$. 
The potential becomes
\begin{equation}
V_{\rm eff}(\sigma,\rho,\theta;T)=
\frac{a(T)}{2}\sigma^2+\frac{\lambda}{4}\sigma^4-H\sigma
-\frac{b_2(T)}{2}\rho^2-\frac{b_3}{3}\rho^3\cos(3\theta)+\frac{b_4}{4}\rho^4
-2h(T)\rho\cos\theta
-g\sigma^2\rho^2 .
\label{eq:Vpolar_wall}
\end{equation}
The wall problem is then a coupled system for $(\sigma,\rho,\theta)$. 
Physically, the dominant topological content is carried by the phase $\theta(z)$, which interpolates between two angular minima separated by $\Delta\theta\simeq\pm 2\pi/3$, while $\rho(z)$ adjusts to reduce the cost of rotating through the anisotropic barrier $\cos(3\theta)$: in the wall core where $\theta'(z)$ is largest, the amplitude $\rho$ typically dips, allowing the trajectory in the complex $\ell$ plane to ``cut the corner'' rather than rotate at fixed radius. 
The chiral field $\sigma(z)$ is then pulled downward (partial chiral restoration) where $\rho$ is large due to the coupling $-g\sigma^2\rho^2$, but can rebound in the core if $\rho$ is suppressed, producing a characteristic three-field structure for the wall energy density.

A controlled analytic handle on the backreaction is obtained by exploiting the hierarchy between the chiral and Polyakov-loop modes. 
In much of the metastability window, the local curvature in the chiral direction,
\begin{equation}
m_\sigma^2(\rho;T)\equiv \frac{\partial^2 V_{\rm eff}}{\partial \sigma^2}
= a(T)+3\lambda\sigma^2-2g\rho^2,
\end{equation}
is parametrically larger than the curvature associated with angular motion in $\theta$. 
In that regime $\sigma(z)$ follows the local minimum determined by the instantaneous $\rho(z)$:
\begin{equation}
\partial_\sigma V_{\rm eff}=0
\quad\Rightarrow\quad
\lambda\sigma^3+\big[a(T)-2g\rho^2(z)\big]\sigma-H=0.
\label{eq:local_sigma_gap}
\end{equation}
Denote the positive solution of \eqref{eq:local_sigma_gap} by $\sigma_\star(\rho;T)$. 
Then, to leading order, one may substitute $\sigma(z)\approx\sigma_\star(\rho(z);T)$ and obtain a reduced functional for the Polyakov loop alone,
\begin{equation}
\mathcal{F}_{\rm red}[\rho,\theta]
\simeq
\int dz\left[
\kappa_\ell\big((\rho')^2+\rho^2(\theta')^2\big)
+V_{\rm red}(\rho,\theta;T)
\right],
\qquad
V_{\rm red}(\rho,\theta;T)\equiv V_{\rm eff}\big(\sigma_\star(\rho;T),\rho,\theta;T\big).
\label{eq:Fred_wall}
\end{equation}
This adiabatic elimination makes the backreaction mechanism transparent: the chiral sector effectively renormalizes the $\rho^2$ term in the Polyakov-loop potential by $-g\sigma_\star^2(\rho;T)$, thereby changing both the preferred radius $\rho_0(T)$ and the stiffness of amplitude fluctuations that govern the wall-core dip. 
Corrections from finite $\kappa_\sigma$ can be included by expanding $\sigma=\sigma_\star+\delta\sigma$; the linearized equation shows that $\delta\sigma$ is sourced by $\rho''$ and $(\theta')^2$ through the $z$-dependence of $\rho(z)$, hence the chiral profile is most strongly distorted where the Polyakov loop rotates fastest.

A practical, physically accurate picture of the wall profile emerges from two complementary approximations. 
First, in the near-$Z(3)$ regime with small $h(T)$, one may treat the wall primarily as an \emph{angular kink} in $\theta(z)$, with $\rho$ and $\sigma$ slaved to $\theta$ through their instantaneous minimization. 
If one further assumes $\rho(z)\approx\rho_0(T)$ approximately constant across the wall (a useful starting point, but not quantitatively exact), then the reduced theory becomes a sine-Gordon-type model in the variable $3\theta$ with a kink connecting $\theta_-$ to $\theta_+$, and the characteristic thickness scale is set by the ratio of angular stiffness $\kappa_\ell\rho_0^2$ to the angular barrier height $\sim b_3\rho_0^3$:
\begin{equation}
\delta_\theta(T)\sim \sqrt{\frac{\kappa_\ell\rho_0^2}{b_3\rho_0^3}}
=\sqrt{\frac{\kappa_\ell}{b_3\rho_0}}.
\label{eq:theta_thickness}
\end{equation}
Second, allowing $\rho(z)$ to vary, one sees directly from the $\rho$-equation in \eqref{eq:Fred_wall} that the ``centrifugal'' term $\kappa_\ell\rho(\theta')^2$ drives $\rho$ downward in the wall core, with the depth of the dip controlled by the amplitude curvature $\partial_\rho^2 V_{\rm red}$ evaluated near the mid-wall angle $\theta\approx(\theta_-+\theta_+)/2$:
\begin{equation}
\delta\rho_{\rm core}
\sim
-\frac{\kappa_\ell\rho_0(\theta')_{\rm max}^2}{\partial_\rho^2 V_{\rm red}(\rho_0,\theta_{\rm mid};T)}.
\label{eq:rho_dip}
\end{equation}
Because chiral backreaction modifies $V_{\rm red}$ through $\sigma_\star(\rho;T)$, it changes $\partial_\rho^2 V_{\rm red}$ and can therefore significantly alter the wall structure. 
In particular, near the spinodal edges identified in Sec.~3, the smallest eigenvalue of the homogeneous Hessian approaches zero, implying that at least one curvature in field space becomes small. 
This generically produces a \emph{thickening} of the wall and a suppression of the surface tension as the barrier flattens. 
Hence, even within a purely mean-field treatment, the domain-wall observables are sharply sensitive to the same metastability/spinodal structure that controls homogeneous decay.

The surface tension $\Sigma(T)$ is defined as the excess free energy per unit area relative to a reference homogeneous vacuum,
\begin{equation}
\Sigma(T)=
\int_{-\infty}^{+\infty}\!dz\;
\left[
\frac{\kappa_\sigma}{2}\sigma'^2
+\kappa_\ell|\ell'|^2
+V_{\rm eff}(\sigma,\ell;T)-V_{\rm ref}(T)
\right],
\label{eq:tension_def}
\end{equation}
where $V_{\rm ref}(T)\equiv V_{\rm eff}(\sigma_{\rm ref},\ell_{\rm ref};T)$ is typically taken as the true-vacuum energy. 
Using the first integral \eqref{eq:firstint_ref}, one obtains an equivalent representation
\begin{equation}
\Sigma(T)=
\int_{-\infty}^{+\infty}\!dz\;
\Big[
\kappa_\sigma\sigma'^2+2\kappa_\ell|\ell'|^2
\Big]
=
2\int_{-\infty}^{+\infty}\!dz\;\Big[V_{\rm eff}(\sigma,\ell;T)-V_{\rm ref}(T)\Big],
\label{eq:tension_alt}
\end{equation}
which is exact for degenerate endpoints and remains an excellent numerical check in the near-degenerate metastable regime. 
In the simplified constant-$\rho$ approximation, $\Sigma$ scales as a kink tension with parametric dependence
\begin{equation}
\Sigma(T)\sim \rho_0^{3/2}\sqrt{\kappa_\ell b_3},
\label{eq:tension_scaling}
\end{equation}
while in the full backreacted system $\rho_0(T)$ and the effective barrier are renormalized by the chiral sector through $g$ and the local solution $\sigma_\star(\rho;T)$. 
This makes $\Sigma(T)$ a sensitive diagnostic of chiral backreaction: holding the Polyakov-loop potential fixed, increasing $g$ typically reduces $\sigma$ in regions of large $\rho$, softens the amplitude curvature, deepens the $\rho$ dip, and tends to reduce the total wall tension, although the competing cost from $\kappa_\sigma(\sigma')^2$ can partially counteract this reduction when $\sigma$ varies sharply. 
In practice, the sign and magnitude of the net effect are extracted by solving the coupled boundary-value problem \eqref{eq:EOMsig_wall}--\eqref{eq:EOMell_wall} and evaluating \eqref{eq:tension_def}.

Numerically, a robust workflow is as follows. 
For each temperature $T$ in the metastability window, determine the relevant endpoint vacua $(\sigma_\pm,\ell_\pm)$ by minimizing $V_{\rm eff}(\sigma,\ell;T)$. 
Initialize a trial profile by interpolating the phase $\theta(z)$ as a tanh-kink from $\theta_-$ to $\theta_+$, choosing a mild Gaussian dip for $\rho(z)$ centered at the wall, and setting $\sigma(z)$ to interpolate between $\sigma_-$ and $\sigma_+$ or to follow the local gap solution $\sigma_\star(\rho(z);T)$. 
Then relax the profile by minimizing $\mathcal{F}$ (gradient flow) or by solving \eqref{eq:EOMsig_wall}--\eqref{eq:EOMell_wall} as a boundary-value problem using collocation; convergence is verified by checking the first integral \eqref{eq:firstint_ref} and the stationarity of $\mathcal{F}$ under small profile variations. 
The wall thickness $\delta(T)$ can be defined operationally as the full width at half maximum of $\theta'(z)$ or of the localized energy density
\begin{equation}
\varepsilon(z)=
\frac{\kappa_\sigma}{2}\sigma'^2+\kappa_\ell|\ell'|^2
+V_{\rm eff}(\sigma,\ell;T)-V_{\rm ref}(T),
\end{equation}
and the surface tension $\Sigma(T)$ follows from \eqref{eq:tension_def}. 
Approaching the spinodal endpoints, one expects $\delta(T)$ to grow and $\Sigma(T)$ to decrease, consistent with the flattening of the effective barrier and the emergence of soft modes; this is the domain-wall manifestation of the same Hessian criterion $\lambda_{\min}(T_{\rm sp})=0$ used in Sec.~3.

\begin{figure}[htp]
  \centering
      \includegraphics[width=0.9\textwidth]{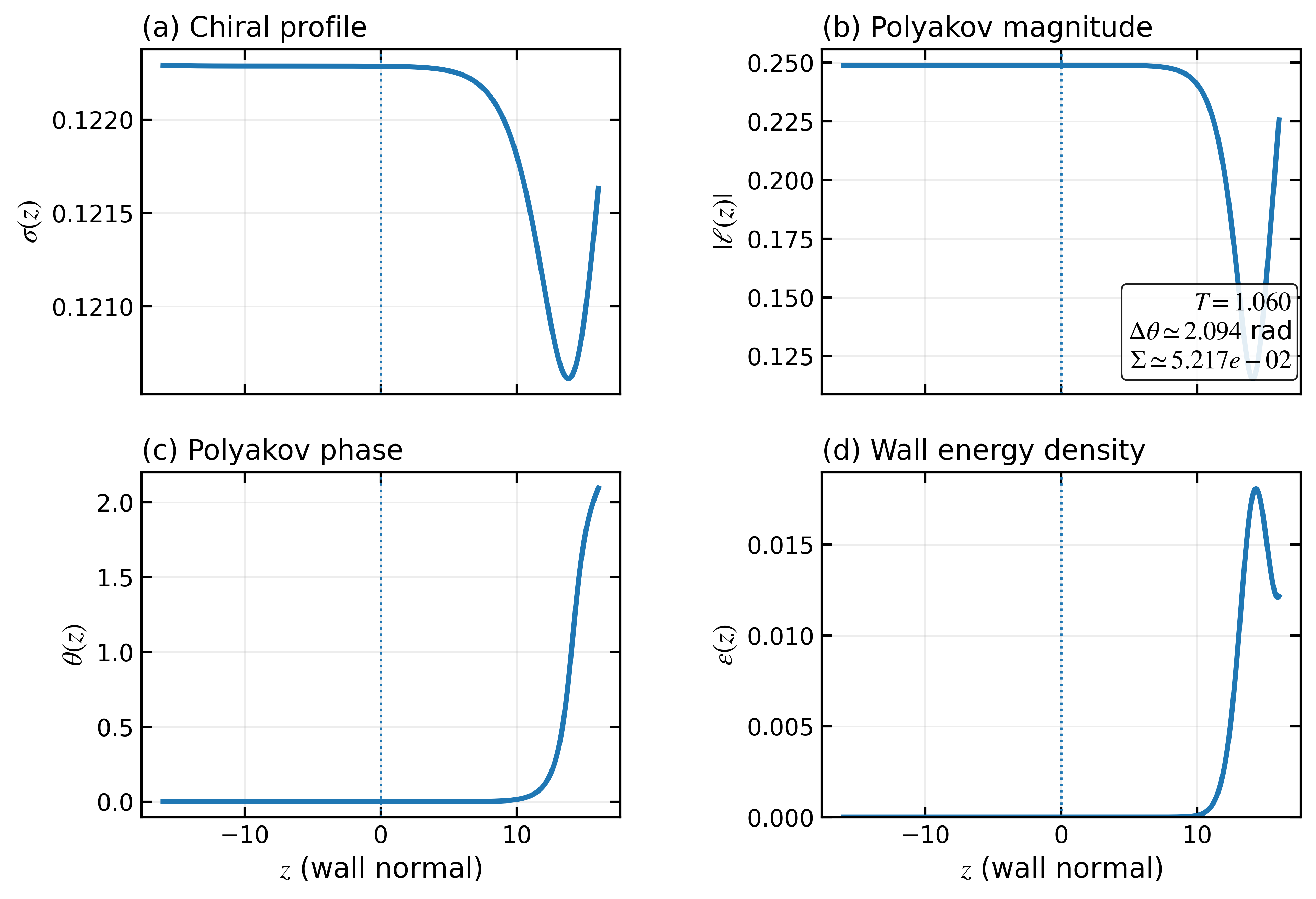}
  \caption{Z(3) domain-wall profiles with chiral backreaction.}
  \label{fig:wall_profiles}
\end{figure}

\begin{figure}[htp]
  \centering
  \includegraphics[width=0.9\textwidth]{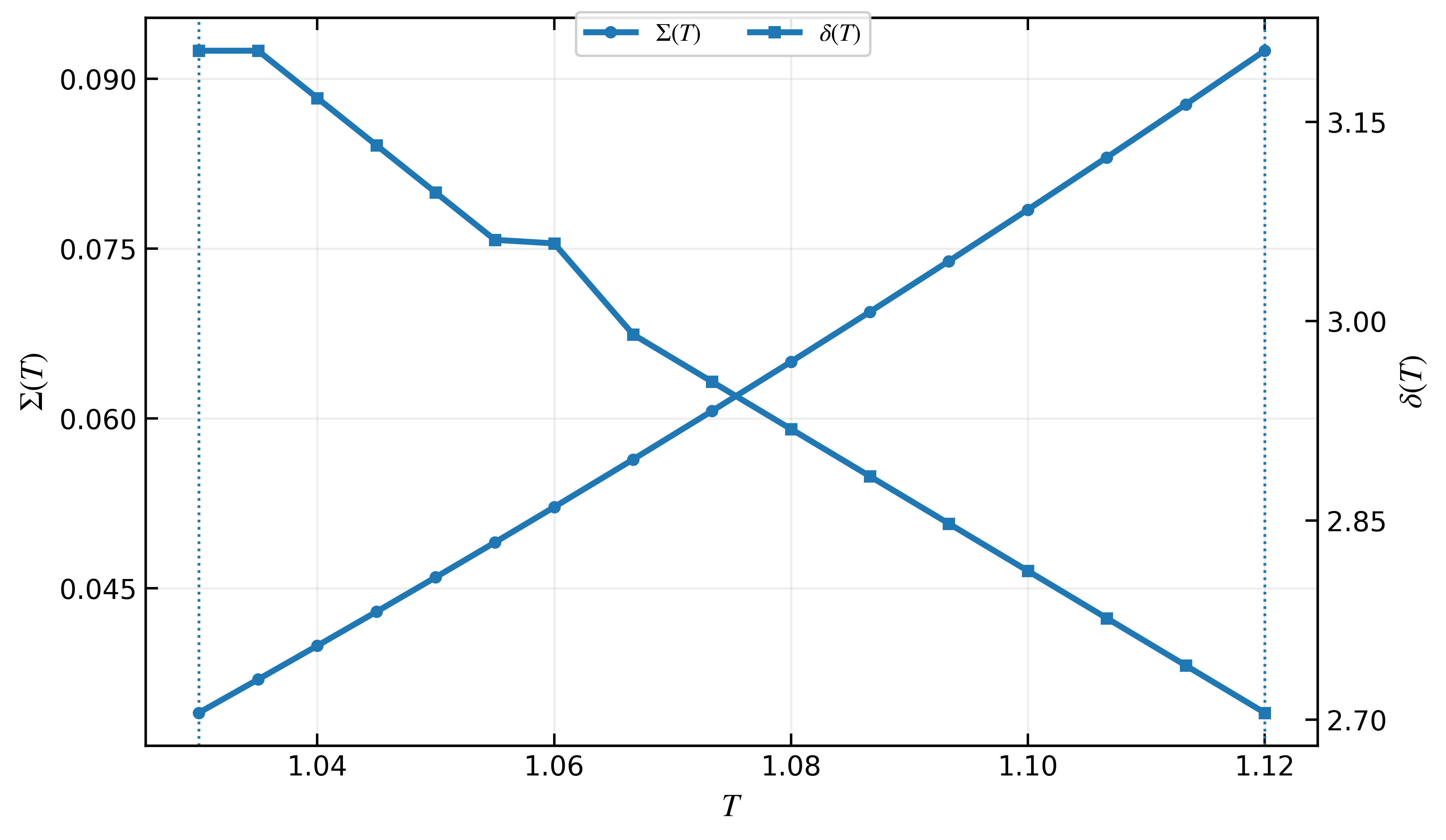}
  \caption{Surface tension and thickness of the Z(3) wall across the metastability window.}
  \label{fig:wall_tension}
\end{figure}

\section{Metastable bubbles and nucleation barrier: $R_c(T)$ and $S_3(T)/T$}
\label{sec:bubbles}

We now turn the equilibrium picture of Secs.~\ref{sec:domainwalls} into a dynamical statement: 
given a metastable (false) homogeneous phase inside the metastability window, what is the dominant escape channel toward the true phase, and how does one compute the critical bubble radius $R_c(T)$ and the nucleation exponent $S_3(T)/T$?
The answer is encoded in the finite-temperature Euclidean bounce, which determines the leading exponential suppression of the nucleation rate per unit volume,
\begin{equation}
\Gamma(T)\;\simeq\;\mathcal{A}(T)\exp\!\left[-\frac{S_3(T)}{T}\right].
\label{eq:Gamma}
\end{equation}
Here $S_3(T)$ is the three-dimensional Euclidean action of the $O(3)$-symmetric saddle, and $\mathcal{A}(T)$ is a prefactor arising from fluctuations around the bounce.
In most theoretical diagnostics the exponential dominates; hence $S_3(T)/T$ is the central quantity that converts the static metastability indicators of Sec.~\ref{sec:homog} into a lifetime statement.
The $O(3)$ nature of the relevant bounce at finite $T$ is a classic result of the dimensional reduction of thermal tunneling \cite{Linde1983,Affleck1981}, and it is the appropriate setting for our coupled chiral--Polyakov effective theory.

Our order-parameter multiplet can be represented as $\Phi\equiv(\sigma,\ell)$ or, equivalently, as real fields $\Phi\equiv(\sigma,\Re\ell,\Im\ell)$, with a temperature-dependent effective potential $V_{\rm eff}(\Phi;T)$ that contains the $Z(3)$ structure in the Polyakov sector and the chiral Landau dynamics, including the backreaction terms fixed in the previous sections.
The metastable state corresponds to a local minimum $\Phi_{\rm false}(T)$, while the stable state corresponds to the global minimum $\Phi_{\rm true}(T)$.
To compute bubble nucleation one introduces the free-energy difference relative to the false vacuum,
\begin{equation}
\Delta V(\Phi;T)\;\equiv\;V_{\rm eff}(\Phi;T)-V_{\rm eff}\!\big(\Phi_{\rm false}(T);T\big),
\qquad \Delta V\!\big(\Phi_{\rm false}(T);T\big)=0,
\label{eq:DeltaV_def}
\end{equation}
and writes the three-dimensional Euclidean action as
\begin{equation}
S_3[\Phi;T]
=
\int d^3x\,
\left[
\frac{1}{2}\,K_{ab}(\Phi;T)\,\nabla\Phi_a\cdot\nabla\Phi_b
+\Delta V(\Phi;T)
\right].
\label{eq:S3_functional}
\end{equation}
The stiffness matrix $K_{ab}$ encodes the gradient cost of spatial inhomogeneities and is the same ingredient that controlled the domain-wall profiles and the surface tension in Sec.~\ref{sec:domainwalls}. 
For the present theoretical treatment we keep the diagonal approximation $K_{\sigma\sigma}=\kappa_\sigma$ and $K_{\ell\ell}=\kappa_\ell$ (with $\ell$ understood as two real components); this is sufficient to obtain controlled scaling relations and to connect nucleation directly to the wall microphysics already computed.

Assuming the dominant saddle is $O(3)$-symmetric, $\Phi(\mathbf{x})=\Phi(r)$ with $r=|\mathbf{x}|$, one obtains
\begin{equation}
S_3[\Phi;T]
=
4\pi\int_0^\infty dr\,r^2
\left[
\frac{1}{2}\,K_{ab}\,\frac{d\Phi_a}{dr}\frac{d\Phi_b}{dr}
+\Delta V(\Phi;T)
\right],
\label{eq:S3_radial}
\end{equation}
and the stationarity condition $\delta S_3=0$ yields the coupled bounce equations
\begin{equation}
K_{ab}\left(\frac{d^2\Phi_b}{dr^2}+\frac{2}{r}\frac{d\Phi_b}{dr}\right)
=
\frac{\partial \Delta V}{\partial\Phi_a},
\label{eq:bounce_multi}
\end{equation}
subject to the regularity and asymptotic boundary conditions
\begin{equation}
\left.\frac{d\Phi_a}{dr}\right|_{r=0}=0,
\qquad 
\lim_{r\to\infty}\Phi_a(r)=\Phi_{a,\rm false}(T).
\label{eq:bounce_bc}
\end{equation}
In a multi-field system the bounce does not simply ``move in one direction''; it chooses an effective path in field space that minimizes the action. 
This is where the coupled chiral--Polyakov structure becomes essential: the system can lower the nucleation barrier by following a valley that combines partial chiral rotation/restoration with Polyakov magnitude and phase rearrangement, rather than paying the full barrier cost in one sector alone. 
This qualitative statement is robust and does not depend on details of the numerical solver.

While a fully general multi-field computation of \eqref{eq:bounce_multi} can be implemented directly (e.g.\ by shooting or relaxation methods in $r$), the most transparent analytic control is obtained in the thin-wall regime, which is relevant near coexistence when the two minima are nearly degenerate and the free-energy splitting between the false and true vacua is small. 
Define the driving force
\begin{equation}
\Delta V(T)\;\equiv\;V_{\rm eff}\!\big(\Phi_{\rm false}(T);T\big)-V_{\rm eff}\!\big(\Phi_{\rm true}(T);T\big)\;>\;0,
\label{eq:DeltaV_T}
\end{equation}
so that $\Delta V(T)$ is the volume free-energy gain per unit volume upon converting false vacuum into true vacuum.
In the thin-wall picture, the free energy of a spherical bubble of radius $R$ can be approximated by a competition between surface cost and volume gain \cite{Coleman1977,CallanColeman1977},
\begin{equation}
F(R;T)\;\simeq\;4\pi R^2\,\Sigma(T)\;-\;\frac{4\pi}{3}R^3\,\Delta V(T).
\label{eq:F_thinwall}
\end{equation}
Here $\Sigma(T)$ is the surface tension of the interface that interpolates between the relevant vacua, and in our framework it is not a free parameter: it is the domain-wall energy per unit area computed from the $Z(3)$ wall profiles with chiral backreaction in Sec.~\ref{sec:domainwalls}. 
Extremizing $F(R;T)$ yields the critical radius,
\begin{equation}
R_c(T)\;=\;\frac{2\,\Sigma(T)}{\Delta V(T)}.
\label{eq:Rc}
\end{equation}
Bubbles with $R<R_c$ shrink because the surface tension dominates, while bubbles with $R>R_c$ grow because the volume driving dominates.
Evaluating \eqref{eq:F_thinwall} at $R=R_c$ gives the thin-wall estimate for the bounce action,
\begin{equation}
S_3(T)\;\simeq\;F(R_c;T)\;=\;\frac{16\pi}{3}\,\frac{\Sigma(T)^3}{\Delta V(T)^2},
\qquad
\frac{S_3(T)}{T}\;\simeq\;\frac{16\pi}{3}\,\frac{\Sigma(T)^3}{T\,\Delta V(T)^2}.
\label{eq:S3_thinwall}
\end{equation}
Equations \eqref{eq:Rc}--\eqref{eq:S3_thinwall} are the key bridge between our earlier sections and the dynamical nucleation picture: $\Delta V(T)$ is determined entirely by the homogeneous vacuum structure (Sec.~\ref{sec:homog}), while $\Sigma(T)$ is determined entirely by the inhomogeneous wall microphysics (Sec.~\ref{sec:domainwalls}). 
Thus, within the same effective theory, the metastability window automatically implies a \emph{nucleation window} whose temperature dependence can be predicted without introducing additional phenomenological inputs.

These formulas immediately imply several robust trends across the metastability window. 
Near coexistence, $\Delta V(T)\to 0$ and therefore $R_c(T)\to\infty$ while $S_3(T)/T$ becomes parametrically large. 
This corresponds to strong exponential suppression and long-lived metastability: the system may remain trapped in the false vacuum for a very long time, and the dominant inhomogeneous structures are then controlled not by bubble nucleation but by pre-existing $Z(3)$ domains and their slow evolution. 
Deeper inside the window, $\Delta V(T)$ increases while $\Sigma(T)$ typically changes more mildly (since it is controlled by stiffness scales and the barrier shape along the optimal wall trajectory), so $R_c(T)$ decreases and nucleation becomes more efficient. 
Finally, approaching the spinodal edge discussed in Sec.~\ref{sec:homog}, the curvature of the metastable minimum softens and the barrier effectively disappears; the decay then crosses over from exponentially suppressed nucleation to a fast instability-driven evolution (spinodal-like decomposition). 
In that regime, the thin-wall formula should be interpreted as a qualitative guide rather than a controlled approximation, but the direction of change remains robust: $S_3(T)/T$ drops rapidly as the spinodal is approached from within the metastable region. 
This provides a direct connection between the metastability indicator (e.g.\ the smallest Hessian eigenvalue of $V_{\rm eff}$ at the metastable minimum) and the dynamical nucleation exponent.

A subtle but important point in a coupled chiral--Polyakov system is that $\Sigma(T)$ and even the effective bubble profile depend on the \emph{field-space trajectory}. 
In a single-scalar theory, the wall is essentially unique; in a multi-field theory, the interface can bend in $(\sigma,\Re\ell,\Im\ell)$ space to minimize its energy. 
A practical way to express this is to parametrize a candidate curve $\Phi_a=\Phi_a(s)$ connecting false and true vacua, and reduce \eqref{eq:S3_radial} to an effective one-dimensional action with an induced metric along the curve,
\begin{equation}
S_3 \;\leadsto\; 4\pi\int dr\,r^2\left[\frac{1}{2}G(s)\left(\frac{ds}{dr}\right)^2+\Delta V_{\rm eff}(s;T)\right],
\qquad
G(s)=K_{ab}\frac{d\Phi_a}{ds}\frac{d\Phi_b}{ds},
\label{eq:path_reduction}
\end{equation}
where $\Delta V_{\rm eff}(s;T)=\Delta V(\Phi(s);T)$. 
The optimal path minimizes $S_3$ over admissible curves. 
Physically, this means that nucleation can proceed by exploiting the softer sector: if the Polyakov phase direction provides a shallow valley near the metastable minimum (approximate $Z(3)$ degeneracy), the bubble can partially rotate $\arg\ell$ while keeping $|\ell|$ and $\sigma$ close to their metastable values, reducing the action cost; conversely, if chiral restoration is already advanced but Polyakov ordering lags, the bounce can proceed primarily in $\sigma$. 
This multi-field flexibility is precisely what makes the dark-QCD chiral transition scenario richer than a single-order-parameter Landau picture, and it is naturally captured by the same effective potential that produced the $Z(3)$ wall backreaction in Sec.~\ref{sec:domainwalls}.

For the present theoretical paper, we therefore adopt a computation workflow that keeps the analysis self-consistent and transparent: 
(i) determine $\Phi_{\rm false}(T)$ and $\Phi_{\rm true}(T)$ and compute $\Delta V(T)$ directly from $V_{\rm eff}(\Phi;T)$; 
(ii) compute $\Sigma(T)$ from the corresponding interface solution (with chiral backreaction) using the same gradient coefficients; 
(iii) use \eqref{eq:Rc} and \eqref{eq:S3_thinwall} as the baseline nucleation diagnostics, and interpret the approach to the spinodal edge using the stability indicator developed in Sec.~\ref{sec:bubbles}. 
This provides a logically closed chain:
\emph{vacuum structure} $\to$ \emph{metastability window} $\to$ \emph{wall microphysics} $\to$ \emph{bubble nucleation barrier}.

\begin{figure}[t]
  \centering
  \includegraphics[width=0.9\textwidth]{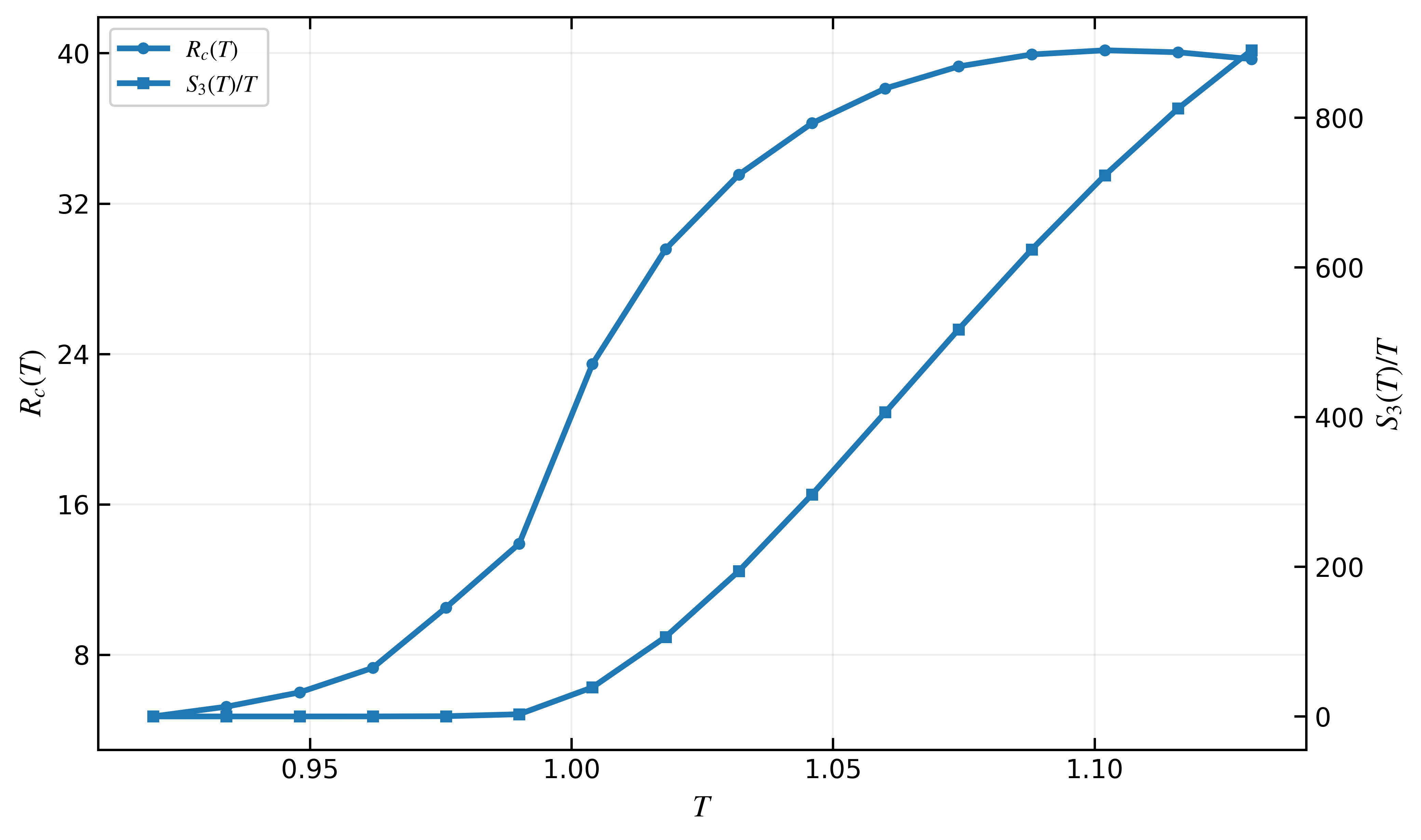}
  \caption{\textbf{Critical bubble radius and nucleation exponent.}
  The critical radius $R_c(T)=2\Sigma(T)/\Delta V(T)$ decreases as the free-energy splitting $\Delta V(T)$ grows away from coexistence, and diverges near coexistence where $\Delta V(T)\to 0$.
  The nucleation exponent $S_3(T)/T \simeq \frac{16\pi}{3}\frac{\Sigma(T)^3}{T\,\Delta V(T)^2}$ is correspondingly large near coexistence (strong exponential suppression) and drops toward the spinodal edge, indicating faster decay of the metastable phase.}
  \label{fig:Rc_S3}
\end{figure}

The $R_c(T)$ and $S_3(T)/T$ provide a compact and model-internal quantification of metastable decay in dark-QCD chiral transitions with a $Z(3)$ Polyakov sector. 
They connect the static potential landscape to a dynamical criterion for whether the system remains trapped (large $S_3/T$), converts by rare nucleation (moderate $S_3/T$), or undergoes a rapid instability-driven decay near the spinodal. 
Because both $\Delta V(T)$ and $\Sigma(T)$ are computed from the same $V_{\rm eff}(\sigma,\ell;T)$ and the same gradient sector used throughout the paper, the nucleation barrier analysis is internally consistent and directly reproducible.

\section{Conclusions}
\label{sec:conclusions}

In this work we developed a fully theoretical framework for a dark-QCD chiral transition in which the confinement-related $Z(3)$ structure of the Polyakov loop is retained and consistently coupled to the chiral order parameter. 
The central goal was to construct a logically closed chain from the homogeneous vacuum landscape to metastability, spinodal loss of stability, and finally to inhomogeneous nonperturbative objects---$Z(3)$ domain walls and metastable bubbles---with explicit, reproducible diagnostics that can be reported as temperature-dependent curves.

Our first main result is a controlled characterization of the \emph{homogeneous} free-energy landscape in the coupled chiral--Polyakov effective theory. 
By tracking multiple $Z(3)$-related extrema as temperature varies, we identified a metastability window in which a false vacuum persists as a local minimum while a true vacuum becomes thermodynamically preferred. 
Within this window the splitting $\Delta V(T)=V_{\rm meta}(T)-V_{\rm true}(T)$ provides a natural measure of how strongly the system is driven away from the false phase. 
Complementarily, the local curvature (Hessian) at the metastable minimum yields a clean \emph{spinodal criterion}: the loss of metastability corresponds to the vanishing of the smallest eigenvalue, signaling that the barrier ceases to exist and the decay mechanism crosses over from exponentially suppressed nucleation to instability-driven evolution.

Our second main result is a quantitatively consistent treatment of \emph{inhomogeneous} $Z(3)$ structures including chiral backreaction. 
We formulated and solved the coupled one-dimensional boundary-value problem for wall profiles connecting distinct $Z(3)$ vacua, allowing the chiral condensate to adjust across the interface. 
This produces two outputs that are simultaneously microscopic (in the sense of being determined by the same effective potential used for the homogeneous analysis) and macroscopic (in the sense of controlling large-scale dynamics): the wall thickness and the surface tension $\Sigma(T)$. 
The backreaction effect is not merely decorative: because the wall trajectory lives in the joint field space $(\sigma,\Re\ell,\Im\ell)$, the energetically preferred interface can bend away from a purely Polyakov-loop interpolation. 
This modifies $\Sigma(T)$ and hence impacts all nucleation-related observables, providing a concrete mechanism by which chiral physics feeds back into confinement-like $Z(3)$ structure in the dark sector.

Our third main result is a nucleation-barrier analysis that connects metastability and wall microphysics in the most direct way. 
Using the thin-wall limit as a transparent baseline, we expressed the critical bubble radius and the thermal tunneling exponent as
\begin{equation}
R_c(T)=\frac{2\Sigma(T)}{\Delta V(T)},\qquad 
\frac{S_3(T)}{T}\simeq \frac{16\pi}{3}\frac{\Sigma(T)^3}{T\,\Delta V(T)^2}.
\end{equation}
These relations make the physics of the metastability window operational: near coexistence, $\Delta V(T)\to 0$ drives $R_c(T)\to\infty$ and yields a large nucleation exponent, implying long-lived trapping; deeper in the window, increasing $\Delta V(T)$ typically reduces $R_c(T)$ and lowers $S_3(T)/T$, making decay more probable; near the spinodal boundary, the rapid softening of the metastable minimum indicates a crossover to fast decay where nucleation is no longer the appropriate language. 
Crucially, since both $\Delta V(T)$ and $\Sigma(T)$ are computed from the same effective theory, the resulting Fig.~6 provides an internally consistent summary of the full chain ``vacua $\to$ metastability $\to$ walls $\to$ bubbles.''

Several natural extensions follow from our analysis. 
First, the thin-wall estimate can be systematically improved by solving the full multi-field $O(3)$ bounce equations, which would quantify deviations when the critical radius is comparable to the wall thickness and would sharpen the approach to the spinodal edge. 
Second, incorporating gauge-sector and fermion-sector parameter variations would allow a more complete mapping of how the metastability window depends on the dark matter content, the dark confinement scale, and explicit $Z(3)$ breaking. 
Third, once the effective theory is embedded into a cosmological history, the temperature-dependent nucleation exponent can be converted into percolation criteria and characteristic length scales of domain formation, opening a route to theoretical predictions for dark-sector inhomogeneities and possible gravitational-wave signatures sourced by first-order dynamics in a hidden confining sector.

In summary, we provided a coherent theoretical pipeline for dark-QCD chiral transitions with explicit $Z(3)$ structure, emphasizing metastability, spinodal criteria, and chiral backreaction on $Z(3)$ domain walls. 
The outputs $\sigma(T)$, $|\ell|(T)$, $\Delta V(T)$, the wall tension $\Sigma(T)$, and the nucleation observables $R_c(T)$ and $S_3(T)/T$ form a compact set of reproducible diagnostics that collectively determine whether the transition proceeds through long-lived trapping, bubble nucleation and growth, or instability-driven decay. 
This framework is designed to be modular and extensible, providing a clean starting point for more refined bounce computations and for connecting dark confinement dynamics to macroscopic phenomenology in hidden-sector cosmology.


\end{document}